# AI Prior Art Search: Semantic Clusters and Evaluation Infrastructure

Boris Genin[1], Alexander Gorbunov[2], Dmitry Zolkin[1], Igor Nekrasov[1]
[1] Division for the Design of Information Search Systems, Federal Institute of Industrial Property, Berezhkovskaya nab. 30-1, Moscow, 125993, Russian Federation
{bguenine, db_dept, igor.nekrasov}@rupto.ru
[2] Development Centre for Artificial Intelligence, Federal Institute of Industrial Property, Berezhkovskaya nab. 30-1, Moscow, 125993, Russian Federation
gorbunov@rupto.ru

**Abstract**

The key to success in automating prior art search in patent research using artificial intelligence (AI) lies in developing large datasets for machine learning (ML) and ensuring their availability. This work is dedicated to providing a comprehensive solution to the problem of creating infrastructure for research in this field, including datasets and tools for calculating search quality criteria.

The paper discusses the concept of semantic clusters of patent documents that determine the state of the art in a given subject, as proposed by the authors. A definition of such semantic clusters is also provided.

Prior art search is presented as the task of identifying elements within a semantic cluster of patent documents in the subject area specified by the document under consideration.

A generator of user-configurable datasets for ML, based on collections of U.S. and Russian patent documents, is described. The dataset generator creates a database of links to documents in semantic clusters. Then, based on user-defined parameters, it forms a dataset of semantic clusters in JSON format for ML.

A collection of publicly available patent documents was created. The collection contains 14 million semantic clusters of US patent documents and 1 million clusters of Russian patent documents.

To evaluate ML outcomes, it is proposed to calculate search quality scores that account for semantic clusters of the documents being searched. To automate the evaluation process, the paper describes a utility developed by the authors for assessing the quality of prior art document search.

**Highlights**

- An infrastructure for researching the use of AI in prior art search is described
- The concept of semantic clusters in patent documents is introduced
- A generator of semantic cluster datasets is described
- A utility for evaluating the quality of prior art searches is described
- Semantic cluster datasets have been created for the US and Russian arrays

## 1. **Introduction**

Advances in the development of AI methods and tools have led to renewed efforts to create systems for automatic prior art search within specified subject areas. The key to success in this endeavor lies in creating and ensuring the availability of large datasets for ML.

Prior art search is the central and most labor-intensive task in patent examination [1,2]. Its quality directly affects the validity of conclusions regarding the patentability of an invention. At the same time, it is widely acknowledged that even experienced experts face a high risk of missing relevant documents due to the vast volume of patent databases and the semantic complexity of technical descriptions [3-5].

Formally, the task of prior art search is to identify all patent documents that characterize the prior art within the subject area specified by the invention application. However, unlike many information retrieval tasks, there is no objective, formalized criterion of relevance: determining which documents truly characterize the prior art remains an intellectual task, requiring expert judgment and a deep understanding of the technical essence [6,7]. These types of tasks do not have known algorithmic solutions, but in recent years, they have been successfully addressed using AI methods, particularly supervised ML [3,8]. Some modern approaches based on deep neural networks and transformers, such as SEARCHFORMER, enable the construction of semantic representations of patent documents and facilitate search based on the similarity of inventions, rather than relying solely on keywords or classification codes [9]. While SEARCHFORMER shows moderate improvements over classical methods, it does not fully resolve the fundamental challenge of patent search: transitioning from textual similarity to an understanding of the technical essence of an invention at a level comparable to that of a human expert.

A key advantage of the patent domain as an AI application area is the availability of large-scale, high-quality training data. Millions of patent publications include official search reports prepared by patent office experts. These reports, along with citations in the INID (56) field, serve as a markup of "correct" search results and can be used to train and evaluate automated search systems without infringing on copyright [10].

This paper proposes a unified approach to solving two interrelated problems:

– creating scalable datasets for ML of prior art search systems;
– evaluating the quality of prior art searches conducted by such systems.

The foundation of the proposed approach is the concept of a semantic cluster of patent documents, a set of documents describing all inventions relevant to a given application. These clusters are built using expert citations and patent families, enabling a shift from document-level to invention-level evaluation of search results.

The paper describes the research infrastructure developed by the authors, which includes:

– a generator of configurable semantic cluster datasets based on US and Russian patent collections;
– a utility for automatic search quality assessment using specialized metrics (*S@K, H@K, MPF@K, MRF@K*) that take into account the structure of semantic clusters.

This infrastructure not only addresses the practical challenge of automating data preparation for training and testing search systems, but also enables objective evaluation of the intelligence level of the AI models used, since successfully solving the state-of-the-art search problem requires understanding the technical essence of inventions, not merely comparing.

## 2. Materials and methods

### *2.1. Automatic creation of datasets for ML*

The first and most important issue concerns datasets for ML. A number of publications [11-16] are dedicated to the creation and application of datasets for ML and the testing of patent search systems.

Recently, new sources have emerged, such as the USPTO's open AI patent dataset, "Artificial Intelligence Patent Dataset (USPTO)" [13], which contains over 13 million patent publications classified by AI technology components. This dataset is intended for training search models and performing novelty analysis, and it includes full texts and metadata.

Patent data analysis is also possible through SQL queries in Google Patents Public Datasets on Google BigQuery [14], which provides access to millions of publications. These resources enable landscape analysis, claim text extraction, and claim breadth estimation.

It should be noted that, to date, researchers and developers of patent search systems have not been provided with approaches or tools to automate the preparation of document collections and datasets.

Patent documentation published by patent offices and WIPO contains a vast amount of useful data, including detailed information on prior art identified by qualified experts and presented in search reports or in the INID (56) field, "Prior art documents", of published patents [10]. This information can be effectively used for the most powerful ML technique: supervised learning. The use of citations in patent documents, especially examination citations, has been noted by a number of authors [17,18]. However, using citations as training data requires consideration of certain specifics. In this paper, the authors describe an approach to the automatic generation of training and test datasets based on expert citations, enriched with data on the patent families of the cited documents.

When creating corresponding datasets for ML, two groups of problems arise. The first group relates to the complexity of processing patent documents, which are mostly presented in XML format according to the WIPO ST.96 standard [10]. Documents in this format contain dozens, or even hundreds, of structured bibliographic fields in addition to the text. However, information about the required prior art search results, obtained by qualified experts, is presented in the form of poorly structured references. Retrieving documents using such references requires specialized expertise in the field of patent information processing. The second group of problems stems from the fact that, during the examination of inventions, it is important for the prior art search results to include information about inventions that may be disclosed in any document belonging to the family of corresponding patents, such as publications in a language familiar to the expert. A patent family is typically defined as a group of documents related to the same invention and includes various publications at different stages of the application life cycle, as well as applications sharing the same set of priority data. In ML, all documents belonging to a corresponding patent family should be included as positive examples. Identifying such documents is a separate task addressed by several well-known patent information systems. In particular, we used the well-known DocDB[1], created and maintained by the European Patent Office (EPO).

For the task of prior art search, the focus is on inventions that define the state of the art in the field of the application under consideration, along with a broad set of patent documents describing these inventions. To represent such a set, the authors introduce the concept of a *semantic cluster* of patent documents [19]. The idea of grouping patent documents based on semantic proximity is not new, cluster-oriented approaches to patent search have been proposed previously [20]. However, as shown below, unlike traditional clustering methods based on textual similarity, the proposed concept of a semantic cluster is grounded in expert citations and patent families.

### *2.2. Semantic clusters of patent documents*

The formal definition of a semantic cluster of patent documents relies on the stable characteristics of patent documents, which stem from the specifics of patent law.

A semantic cluster of patent documents is defined as a set that includes:

- the invention application, the corresponding patent description (if a patent is granted), and other publications containing full-text information at various stages of the invention's life cycle;

---

[1] https://www.epo.org/en/searching-for-patents/data/bulk-data-sets/docdb

- patent documents from the family of corresponding patents that include the invention application (including all publications containing full-text information at various stages of the invention's life cycle)

- patent documents cited by experts in publications resulting from the examination, including search reports and the INID (56) field in the grant publication;

- patent documents from the families of corresponding patents that include the documents cited by the expert

Thus, the semantic cluster of a base document $x$ is defined as the combination of the patent family of document $x$ $(Cx)$ and all patent families $Cy$, where $y \in Yx$, and $Yx$ is the set of documents cited by the examiner in the prior art search report for document $x$. In other words, the semantic cluster consists of documents describing inventions that, in the examiner's opinion, should be taken into account when examining the application under consideration. Figure 1 illustrates the structure of the semantic cluster for a typical U.S. patent document.

Given this definition, the solution to the problem of automatic prior art search, for the purposes of examining a pending invention application, is to retrieve documents that belong to the corresponding semantic cluster.

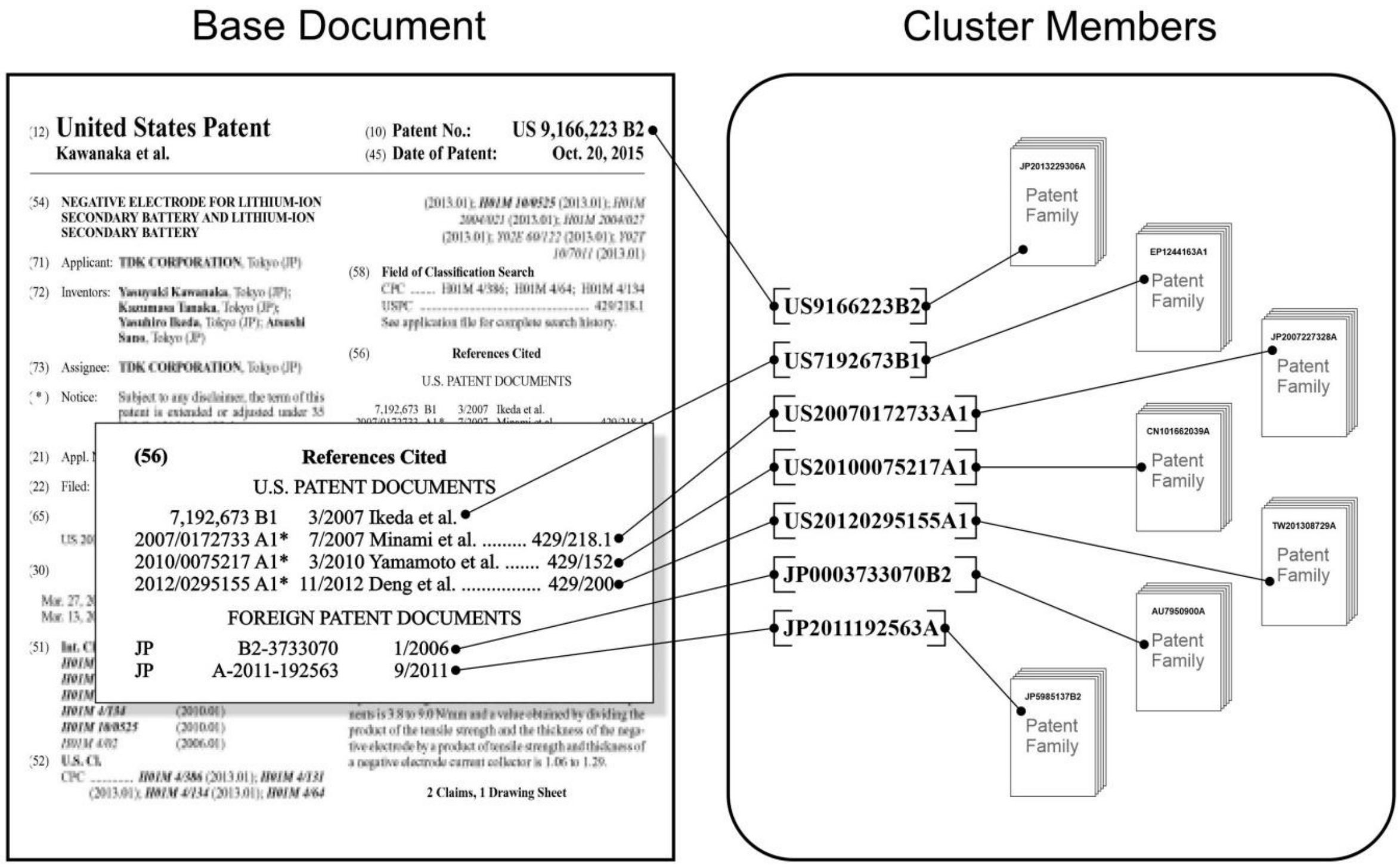

**Fig. 1.** Example of a semantic cluster for patent US 9166223 B2 (base document of the cluster).

*2.3. Semantic cluster dataset generator*

The dataset generator identifies semantic clusters of patent documents and creates a database containing references to the documents included in these clusters [21]. Based on user-specified parameters, a set of labeled data for ML is then generated. The generator can also be used to produce a test set for evaluating ML results by calculating the search quality score of the tested system using a dedicated utility.

The system for generating semantic cluster datasets is based on representing clusters as an SQL database managed by the PostgreSQL database management system.

Each entry in this database contains references to the identifiers of documents belonging to a single cluster, including:

- Base document identifier (<document office code> <document number> <document type code>_< document publication date>);
- Identifiers of patent family members of the base document;
- Identifiers of references cited in the base document to patent documents from the same patent office as the base document, as well as patent family members of these documents;
- Identifiers of references cited in the base document to patent documents from a patent office different from that of the base document, along with the patent family members of these documents.

This structure makes it possible to use for ML a dataset of semantic clusters containing patent documents in JSON format, generated "on the fly".

The size of the dataset can be adjusted by specifying a publication date range.

To create thematic datasets, one can use lists of documents obtained by searching either through classification data or any thematic queries supported by the system.

Semantic clusters can be generated by the described generator from the patent collections of various patent offices, or from combinations of them. The results of generating semantic cluster collections from the U.S. and Russian patent collections are described below.

*2.4. Datasets of semantic clusters for US and Russian patent collections*

The collection of semantic clusters of U.S. patent documents was formed from publicly available publications of U.S. patent documents.

The collection includes documents with publication dates from 2001 to June 2024: specifically, applications with the document type code A1 (Utility Patent Application published on or after January 2, 2001) and granted patents with the code B2 (Utility Patent Grant with pre-grant publication, issued on or after January 2, 2001).

Some characteristics and features of the created collection of semantic clusters of U.S. patent documents, as shown in Table 1, are of particular interest.

**Table 1** General characteristics of the collection of semantic clusters of US patent documents

| Indicator | Application | Patents | Total |
|---|---|---|---|
| Number of clusters (database entries) | 9,391,676 | 5,024,240 | 14,415,916 |
| Number of non-unique documents in all clusters | 168,638,211 | 588,767,602 | 757,405,813 |
| Number of unique documents in all clusters | | | >30,000,000 |
| Number of clusters containing only the base document | 655,619 | 11,040 | 666,659 |
| Number of citations from the base patent office | 12,320,745 | 113,842,829 | 126,163,574 |
| Number of citations from other patent offices | 38,294 | 543,028 | 581,322 |
| Clusters containing citations from the base patent office | 1,388,169 | 4,592,763 | 5,980,932 |
| Clusters containing citations ONLY from the base patent office | 1,364,697 | 4,512,557 | 5,877,254 |
| Clusters containing citations from NOT the base patent office | 25,906 | 82,103 | 108,009 |
| Clusters containing citations ONLY NOT from the base patent office | 2,434 | 1,897 | 4,331 |
| Clusters containing citations BOTH from the base office and non-base office | 23,472 | 80,206 | 103,678 |

| Number of clusters with only patent family members of the base document (without citations) | 7,345,454 | 418,540 | 7,763,994 |
|---|---|---|---|
| Average number of all citations | 1.32 | 22.77 | 8.79 |
| Average number of citations from the base patent office | 1.31 | 22.66 | 8.75 |
| Average number of citations from other patent offices | 0.00 | 0.11 | 0.04 |
| Average number of patent family members for the base document | 7.46 | 8.66 | 7.88 |

Figure 2 shows the distribution of the number of clusters for US patents by the number of documents contained in each cluster.

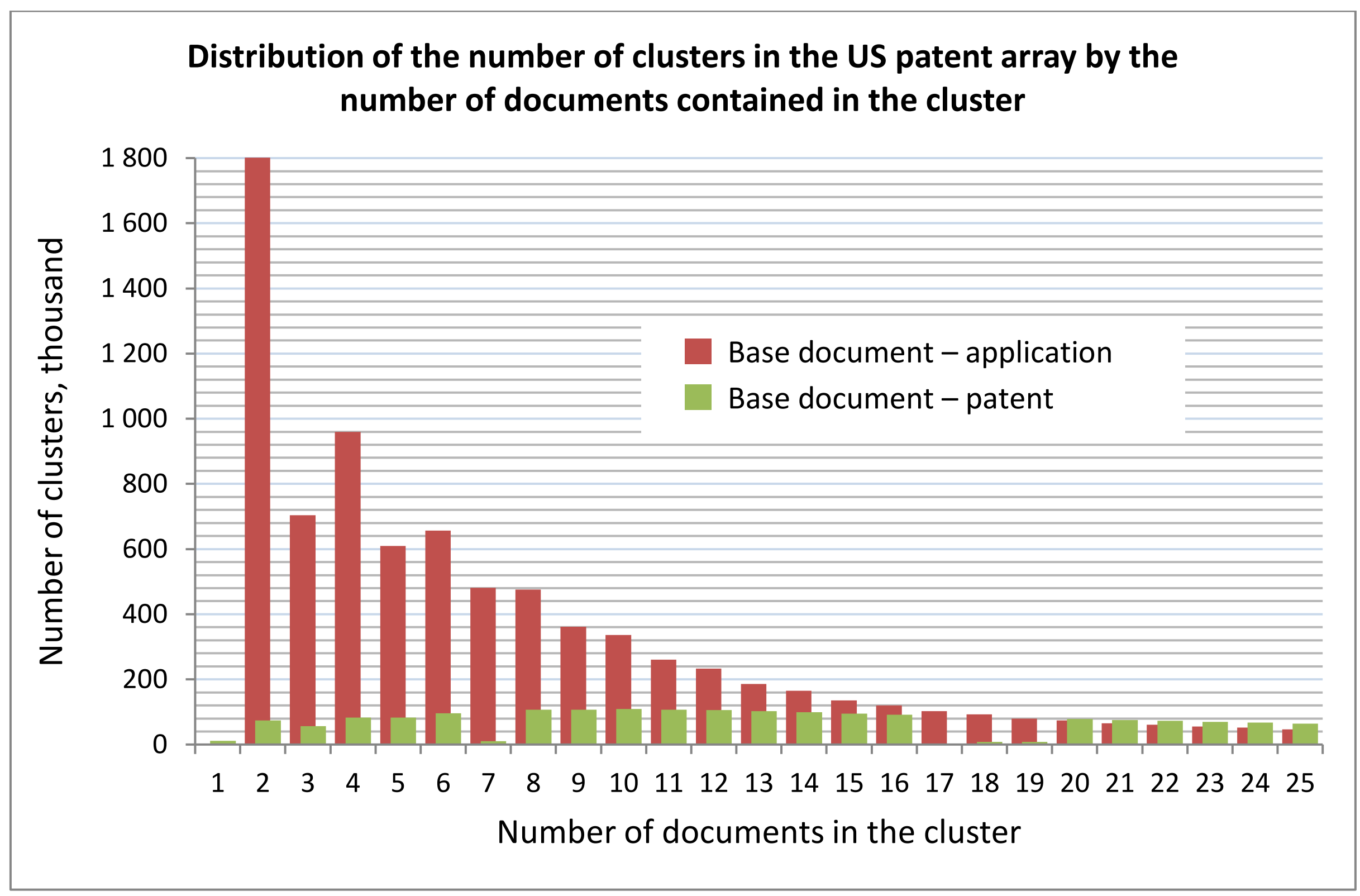

**Fig. 2.** Distribution of the number of clusters in the US patent array by the number of documents contained in each cluster.

The graph shows that approximately 960,000 application-based clusters and 80,000 patent-based clusters each contain only four documents. It can also be seen that application-based clusters and about 85,000 patent-based clusters each contain 15 documents. The number of "large" clusters gradually decreases as the number of cluster members increases. These characteristics are essential for ML using semantic cluster datasets.

Table 1 presents some general features of the U.S. patent document dataset identified during the formation of semantic clusters. In particular, two features of the examination of invention applications at the U.S. Patent and Trademark Office may be especially relevant to our task. First, a large number of search reports contain extensive lists of documents characterizing the state of the art in the subject area of the invention application. Second, most of the documents included in the examiner's prior art assessment are U.S. publications.

The presented implementation of the collection of semantic clusters of Russian patent documents was formed from publicly available publications with publication dates from 2001 to July 2025. This includes applications with document type code A and patents with codes C1 and C2. The main characteristics and features of the created collection of semantic clusters of Russian patent documents are shown in Table 2.

**Table 2** General characteristics of the collection of semantic clusters of Russian patent documents

| Indicator | Applications | Patents | Total |
|---|---|---|---|
| Number of clusters (database entries) | 462,720 | 649,782 | 1,112,502 |
| Number of non-unique documents in all clusters | 3,900,202 | 11,988,273 | 15,888,475 |
| Number of unique documents in all clusters | 3,046,751 | 5,969,911 | 6,730,021 |
| Number of clusters containing only the base document | 160,444 | 90,618 | 251,062 |
| Number of citations from the base patent office | 0 | 790,111 | 790,111 |
| Number of citations from other patent offices | 0 | 581,840 | 581,840 |
| Clusters containing citations from the base patent office | 0 | 408,720 | 408,720 |
| Clusters containing citations ONLY from the base patent office | 0 | 192,358 | 192,358 |
| Clusters containing citations from NOT the base patent office | 0 | 298,768 | 298,768 |
| Clusters containing citations ONLY NOT from the base patent office | 0 | 82,406 | 82,406 |
| Clusters containing citations BOTH from the base office and non-base office | 0 | 216,362 | 216,362 |
| Number of clusters with only patent family members of the base document (without citations) | 302,276 | 68,038 | 370,314 |
| Average number of all citations | 0.00 | 2.11 | 1.23 |
| Average number of citations from the base patent office | 0.00 | 1.22 | 0.71 |
| Average number of citations from other patent offices | 0.00 | 0.90 | 0.52 |
| Average number of patent family members for the base document | 7.43 | 5.28 | 6.18 |

Figure 3 shows the distribution of the number of clusters for Russian patents by the number of documents contained in the cluster.

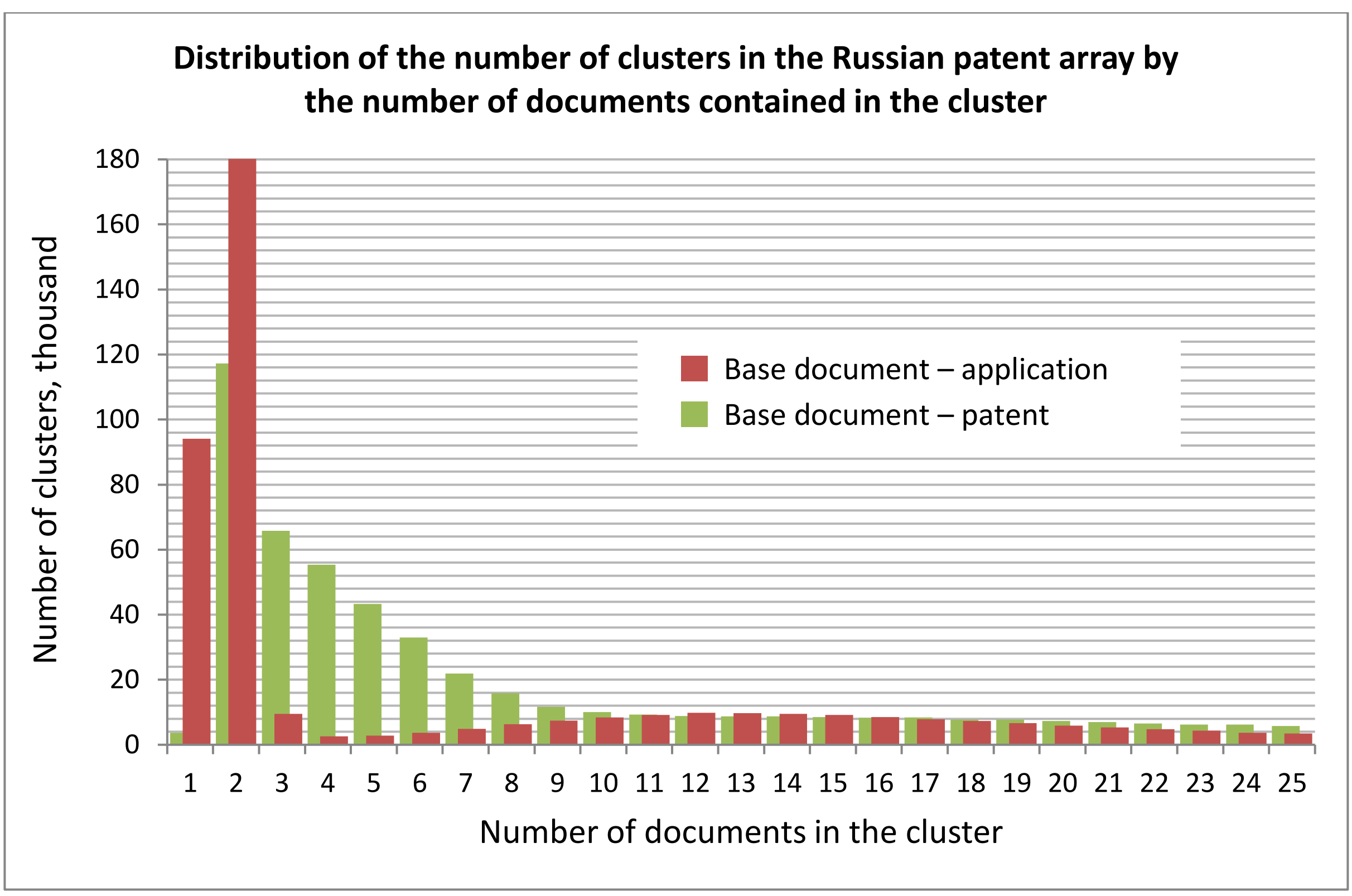

**Fig. 3.** Distribution of the number of clusters in the Russian patent array by the number of documents contained in the cluster.

Table 2 and the graph in Figure 3 illustrate the specific characteristics of the array of semantic clusters of Russian patent documents. A notable feature of this dataset is the publication of the search report for an application as a separate document, typically in an unstructured format. As a result, search report data are not included in the publication of the application itself.

*2.5. Method and utility for evaluating the quality of automatic prior art search*

*2.5.1. Quality criterion for automatic patent search*

A patent document identified during an automatic prior art search is considered relevant if it belongs to the semantic cluster that includes the application for which the search is being conducted. However, in patent search, the objective is not merely to retrieve documents but to identify the inventions they describe, those that characterize the prior art relevant to the application under consideration. This implies that identifying multiple documents from the same patent family should not increase the search quality score. The "unit of relevance" in a patent search is not the individual patent document but the patent family. In other words, if two relevant documents from the same family, describing the same invention, are retrieved, only one relevant result is counted. Moreover, documents from the patent family of the application being evaluated should not be considered relevant when calculating search quality scores.

*2.5.2. Evaluating the quality of automatic prior art search*

Common indicators used to evaluate search quality include precision and recall [22] as well as other indicators based on precision and recall: precision@K/recall@K, F-score/F1-score, Average Precision (AP), Mean Average Precision (MAP), etc. [23]. In addition, metrics that account for both the relevance and the position of documents in the search results are widely used, such as R-precision [22,24], DCG, nDCG, α-nDCG [23-26]. Most of these metrics are designed to assess search result quality for a single query (with MAP being a notable exception). In the simplest case, the performance of a search system over a set of queries is evaluated by averaging the scores for each individual search. However, the varying number of relevant documents per query and the spread of relevance ratings across queries necessitate increasingly sophisticated approaches to evaluating performance across large-scale systems. For example, the TREC conference uses Mean-nDCG@10, which averages normalized DCG for the top 10 results and takes into account the distribution of nDCG across queries, the median, and statistical significance [27]. Google also uses Mean-nDCG@10, along with Mean Reciprocal Rank (referred to as “Mean MRR” in the source), and segments queries to reduce the influence of high-frequency queries on system performance [28].

An important limitation of these traditional search quality assessments, when applied to patent search, is their focus on document-to-document comparison, that is, evaluating the system’s ability to find a *document* relevant to a given document. However, in patent search, the objective is to identify relevant *inventions*, which are typically represented by a group of documents belonging to the same patent family.

The idea of evaluating search quality not by individual documents but by groups of relevant documents (e.g., topics, entities, or clusters) is not new. In the information retrieval literature, several metrics have been proposed to address this, taking into account diversity and subtopic coverage. These include as α-nDCG [26], Subtopic Precision and Recall [29] and Intent-Aware metrics [30]. However, these approaches often require manual or semi-automatic labeling of documents by category or rely on a probabilistic model for determining user intent.

The described approach evaluates the system’s ability to retrieve any documents representing patent families from a relevant cluster (i.e., relevant patent families), assuming that all documents within a cluster are equally relevant. At the same time, as shown in Tables 1 and 2, cluster sizes can vary significantly, as can the number of examination references included in the search report. To address this, the authors propose two performance measures: (1) the average number of searches in which at least one relevant patent family was found, and (2) the average number of searches in which the system successfully retrieved all relevant patent families.

Let:

$K$ – the specified number of values considered in the ranked list of search results, $K \in \mathbb{N}_1$(e.g. $K$=20);

$R_K$ – the set of the first $K$ documents *(top (K))* found by the system;

$x$ – the document for which the prior art search is performed, i.e. document under consideration;

$C_x$ – the patent family for document $x$, $|C_x| \geq 1$. Documents from this patent family, if they appear in the search results, are not considered relevant;

$Y_x$ – the set of documents referenced by the expert in the search report for document $x$;

For each document $y \in Y_x$, there is a patent family $C_y$, which includes this document: $|C_y| \geq 1$;

Then the set of unique patent families for documents cited by the expert will be defined as: $C_Y=\{C_y/y \in Y_x\}$ (each family is included only once).

From the definition, it follows that the union of $C_x$ and $C_Y$ is the semantic cluster of document $x$;

$N$ – the number of searches performed.

Then, the quality assessment of the automatic prior art patent search is defined as the ratio of the number of searches in which at least one relevant document appears in $R_K$ to the total number of searches. In this case, the probability of finding at least one relevant document (at least one representative of at least one relevant patent family) in each search $S@K$ (Success@K) will be defined as:

$$S@K = \frac{1}{N}\sum_{i=1}^{N}(\pi_k), \tag{1}$$

where:

$(\pi_k)$ – metric calculated using the formula:

$$(\pi_k) = \begin{cases} 1, \text{if } R_K \cap C_Y \neq \emptyset, \\ 0, \text{otherwise} \end{cases} \quad (2)$$

This metric is similar to the hit rate (Hit@K), a commonly used metric in a number of applied tasks [31,32]. It is typically employed in systems where it is important to determine whether a relevant item appears among the *top (K)* results, for example, in search engines where users are unlikely to review all results.

The $H@K$ (Hit-all-possible@K) metric estimates the proportion of searches in which the system, for each document in the search results:

- finds all relevant families if the expert has specified fewer than *K* links in the search report;
- finds *K* different relevant families (one document from each) if the expert specified at least *K* links in the search report. In this case, the set $R_K$ must consist entirely of documents belonging to unique patent families.

$$H@K = \frac{1}{N}\sum_{i=1}^{N}(\rho_{Ki}), \quad (3)$$

where for each $C_y \in C_Y$ search results in the *i*-th search ($i$ = 1..*N*), the metric $c_y$ is calculated – the indicator of search success for the patent family:

$$c_y = \begin{cases} 1, \text{if } R_K \cap C_y \neq \emptyset \\ \text{otherwise } 0 \end{cases}; \quad (4)$$

for each *i*-th search, the metric $(\rho_K)$ is calculated – the indicator of search success for the cluster:

$$(\rho_K) = \prod_{y=1}^{Y}(c_y). \quad (5)$$

$H@K$ can be interpreted as the probability of including the maximum possible number of documents from relevant patent families within $R_K$ for each search.

The advantages of $S@K$ and $H@K$ include the clarity of evaluation results from the perspective of solving the problem as a whole.

The known estimates of average precision and recall can be refined to take into account patent families and can be represented as mean precision_family@K (*MPF@K)* and mean recall_family@K (*MRF@K*), respectively:

$$MPF@K = \frac{1}{N}\sum_{i=1}^{N}(PF@K_i); \quad (6)$$

$$MRF@K = \frac{1}{N}\sum_{i=1}^{N}(RF@K_i). \quad (7)$$

Here:

$PF@K = \frac{|\{C_y \in C_Y | R_K \cap C_y \neq \emptyset\}|}{K}$ – the precision of a specific search (precision_family@K, *PF@K*), that is, the ratio of the number of unique documents belonging to relevant patent families found as a result of searching the *top (K)* documents in the search results list to the total number of documents returned by the system, i.e. *K*.

$RF@K = \frac{|\{C_y \in C_Y | R_K \cap C_y \neq \emptyset\}|}{|C_Y|}$ – the recall of a specific search (recall_family@K, *RF@K*), that is, the ratio of the number of unique documents belonging to relevant patent families found as a result of searching the *top (K)* documents in the search results list to the total number of documents corresponding to the query, or more precisely, to the number of references in the search report (the size of the set $C_Y$).

We believe that the proposed metrics are better suited for evaluating the performance of automatic patent search systems than metrics traditionally used in general information retrieval tasks. For example, calculating metrics such as *MAP* and *nDCG*, popular in information retrieval, requires determining relevance (binary or graded) *at the document level*. However, the core of the proposed concept is that relevance should be assessed at the *patent family* (i.e., invention) level. Therefore, using *MAP/nDCG* to evaluate searches based on semantic clusters does not account for the specifics of the problem statement. These metrics do not capture the value that is critical for patent examination. They measure document-level precision and recall, whereas in patent search, what matters is semantic (invention-level) precision and recall.

In contrast, the proposed metrics, $S@K$, $H@K$, $MPF@K$, and $MRF@K$ account for the specific nature of patent search. They are interpretable, focused on the actual goal (identifying inventions rather than individual documents), and provide a practically useful representation of system effectiveness.

*2.5.3. Search quality assessment utility*

As noted above, datasets based on semantic clusters can be used to assess the quality of automatic prior art searches and to conduct comparative analyses of patent search systems. As part of the effort to create a research infrastructure for patent search automation, a utility for evaluating the quality of automatic patent searches was developed. The utility consists of three main blocks.

The first block performs the search and selection of documents for testing using the API of Rospatent's patent search platform. To initiate a search, it is necessary to specify either a specific publication date or a range of dates, along with one or more document type codes corresponding to registered (granted) patents. To further reduce the test dataset, it is possible to select every second (or third, etc.) document from the list of search results based on the chosen criteria; this sampling interval is defined by a parameter. At the final stage of selection, the utility checks for the presence of active patent references (which are required for calculating the search quality metrics), as well as the presence of at least one of the text fields from the invention description: abstract, description, or claims. This block outputs (and saves) a list of the identifiers of the selected documents and their total number.

The selection process may be skipped if a previously created and saved list of document identifiers is to be used for the search.

The utility allows the transfer of the list of selected document identifiers to any system performing a prior art search and enables the retrieval of search results from that system. Searches for each selected document can also be performed directly within the global patent collection on the Rospatent Patent Search Platform, using the platform's API.

The result of this block is a set of search results for the selected sample documents, sorted by relevance. The results list includes the *top (K)* search results.

The next block calculates the previously described prior art search quality metrics – $S@K$, $H@K$, $MPF@K$, and $MRF@K$ based on the search results obtained from the tested system. These metrics are calculated by determining whether the retrieved documents belong to the relevant patent families included in the corresponding semantic cluster.

As a result, the utility outputs the calculated quality assessment values for the automatic patent search and saves the search results, along with the corresponding relevance indicators used in the calculations, in output files.

## 3. Discussion

The infrastructure proposed in this paper for generating datasets and evaluating the quality of patent search enables the creation of semantic cluster datasets containing patent documents that an automatic prior art search system should identify and take into account, with minimal effort and without the need for costly expert involvement. Importantly, the proposed infrastructure includes a utility that, when evaluating automatic search systems, considers the multiplicity of documents describing an invention present within the generated datasets.

A possible extension of this approach would be to enrich the datasets with non-patent literature, which is essential in several technological fields where experts frequently cite scientific publications in prior art search reports.

The research infrastructure proposed in the paper, including a dataset generator based on semantic clusters and an evaluation utility with *S@K, H@K, MPF@K, and MRF@K* metrics, offers a new opportunity to objectively assess the intelligence level of the AI models used.

Indeed, prior art search is an intellectual task that lacks a clear algorithm. When compiling a search report, experts do not follow strict rules; instead, they analyze the essence of an application, search for analogues, compare features, and assess novelty and inventive step.

Therefore, we propose considering the success in solving the prior art search task as a measure of a system's intelligence.

Thus, the developed infrastructure extends beyond serving as a tool for addressing a highly specialized problem. It becomes a test environment for evaluating AGI (Artificial General Intelligence) in an applied, measurable context. The paper proposes a specific and quantifiable criterion for assessing how well AI can handle a task that requires understanding the essence of an invention.

## 4. Conclusion

The task of automatic prior art search for the purpose of examining patent applications has been considered. It is proposed to define the solution to this task as maximizing the proximity of the results of an automatic patent search to those of an expert search, as reflected in the search report accompanying the patent application.

The concept of a “semantic cluster of patent documents” has been introduced. Such a cluster includes patent documents that describe inventions constituting the state of the art within a specific subject area.

The implementation of a generator for semantic cluster datasets of U.S. and Russian patent documents, intended for ML and the evaluation of patent search systems, has been described. This implementation includes databases comprising 14.4 million semantic clusters (over 757 million non-unique U.S. patent documents) and 1 million semantic clusters (over 15 million non-unique Russian patent documents).

Criteria have been proposed for evaluating the quality of automatic prior art searches, taking into account the representation of documents describing different inventions and belonging to various corresponding patent families.

A utility for evaluating the quality of automatic prior art search systems has been described. It uses a test dataset of semantic clusters of patent documents defined by the researcher. The developed research infrastructure for automatic patent search using AI is freely available on the Rospatent search platform.

## CRediT authorship contribution statement

Boris Genin: Conceptualization, Methodology, Validation, Investigation, Writing – original draft, Writing – review & editing, Visualization.

Alexander Gorbunov: Conceptualization, Methodology, Formal analysis, Writing – original draft, Writing – review & editing, Supervision, Funding acquisition.

Dmitry Zolkin: Investigation, Project administration, Funding acquisition, Resources.

Igor Nekrasov: Software, Validation, Visualization.

## Funding

The research did not receive any specific grant from funding agencies in the public, commercial, or not-for-profit sectors.

## Declaration of competing interest

The authors declare no conflict of interest.